# Nonlocal Optical Response of Particle Plasmons in Single Gold Nanorods

*Weixiang Ye[1], ****

[1] Center for Theoretical Physics and School of Science, Hainan University, Haikou 570228, China

*Corresponding author: wxy@hainanu.edu.cn

ABSTRACT:

Particle plasmons in metal nanoparticles have primarily been investigated through the use of local optical response approximations. However, as nanoparticle size approaches the average distance of electrons to the metal surface, mesoscopic effects such as size-dependent plasmon linewidth broadening and resonance energy blue shifts are expected to become observable. In this work, we compared the experimental spectral characteristics with simulated values obtained using a generalized nonlocal optical response theory-based local analogue model. Our results show that the nonlocal plasmon damping effects in single nanoparticles are less significant compared to those observed in plasmon-coupled systems. Moreover, our study demonstrates that single-particle dark-field spectroscopy is an effective tool for investigating the nonlocal optical response of particle plasmons in single nanoparticles. These results have important implications for the rational design of novel nanophotonic devices.

Metal nanoparticles have revolutionized the field of nanophotonics by enabling light concentration on length scales smaller the diffraction limit through particle plasmon excitation.[1] This unique optical property has opened up numerous new applications in the field of nanophotonics, such as nanolasers, optical metamaterials, and biochemical sensing on the nanoscale. [2-7] Investigating particle plasmons at the single-particle level is crucial to understand the physical mechanisms behind these applications. Single-particle dark-field spectroscopy is a highly effective tool for single-particle characterization due to its ease of implementation, high-throughput capabilities, and superior spectroscopic resolution. Using this technique, spectral features, such as resonance energy $E_{res}$, line width $\Gamma$, and scattering intensity $I_{sca}$ of a single metal nanoparticle can be obtained, providing insights into the particle's shape, material composition, surface charges, and local environment.[8-10]

Recent studies have revealed that the confinement of electrons within small nanoparticles promotes nonlocal optical response, resulting in significant blue shifts of the $E_{res}$ and broadening of the $\Gamma$.[11-13] This phenomenon has been explored both experimentally and theoretically in single silver nanoparticles using electron energy loss spectroscopy (EELS).[14,15] Although optical measurements have been applied to investigate the nonlocal optical response of Au film-coupled nanoparticles, superlattice monolayer of gold nanoparticles, and ultrathin metal-dielectric-metal planar structures, [16-18] investigation of nonlocal optical response in single gold nanoparticles remains elusive. The difficulty arises from the relatively large inter-band damping of gold nanoparticles in the visible range, which weakens the detection of the nonlocal optical response.[19]

In this Letter, we present a theoretical and experimental investigation of nonlocal optical response in single gold nanorods using single-particle dark-field spectroscopy. By exploiting the tunable scattering spectra of gold nanorods via their aspect ratio, we investigate the nonlocal optical response of gold nanorods in the near-infrared region where inter-band damping is absent. To examine the role of induced charge diffusion constant ($\mathcal{D}$) and electron convection constant ($\beta$) in determining the nonlocal optical response of single gold nanorods, we compare experimentally obtained spectral characteristics with values simulated by the boundary element method (BEM) using a generalized nonlocal optical response theory-based local analogue model (GNOR-LAM). [12,20-22] We find that the value of $\mathcal{D}$ is smaller than the previously reported experimental value in the plasmon-coupled system. Our method provides a promising approach to comprehending nonlocal optical responses on a single nanoparticle level. We anticipate that our method could help unify electron-surface scattering damping, chemical interface damping, and Landau damping in particle plasmon as they may all arise from induced charge diffusion effects. Furthermore, we observe that $\beta$ conforms quite well to the relationship $\beta = \sqrt{3/5 v_F^2}$ (where $v_F$ is the Fermi velocity)

In the plasmonic nanoparticle research field, the local-response approximation (LRA) is a commonly used approach for the theoretical modeling of the optical response of nanoparticles. Although this method has successfully explained various plasmonic phenomena, including electron energy-loss spectroscopy, cathodoluminescence experiments, near-field microscopy, and optical far-field measurements, it faces limitations such as size-dependent plasmon linewidth broadening and blue shift of the plasmon resonance energy.[23] To address these issues, Mortensen *et al*. proposed a semiclassical generalized non-local optical response theory (GNOR) that unites quantum pressure convection effects and induced charge diffusion kinetics. The GNOR theory has successfully explained both size-dependent plasmon linewidth broadening and resonance energy blue shifts in single metallic nanoparticles.[12] To facilitate its implementation in any electromagnetic simulation platform, we have extended it to the Generalized nonlocal optical response theory-based local analogue model (GNOR-LAM) by combining the local analogue model (LAM) with GNOR theory.[21-22] This approach eliminates the need for implementing a *k*-dependent permittivity and simplifies the simulation of plasmonic nanoparticles of different shapes. In the frequency domain, the optical response of metals can be described using the coupled electromagnetic equations with GNOR theory:[23]

$$\nabla \times \nabla \times E(r,\omega) = (\omega/c)^2 \varepsilon_{core}(\omega) E(r,\omega) + i\omega\mu_0 J(r,\omega) \quad (1)$$

$$[\beta^2/\omega(\omega+i\gamma) + \mathcal{D}/i\omega]\nabla[\nabla \cdot J(r,\omega)] + J(r,\omega) = \sigma(\omega) J(r,\omega) \quad (2)$$

Here, the induced current density $J(r,\omega)$ represents the movement of free electrons in response to the incident electromagnetic field, while the dielectric response from the bound electrons is denoted by $\varepsilon_{core}(\omega)$. In addition, $\sigma(\omega)$ and $\gamma$ are the Drude conductivity and damping rate of the metals, respectively. The induced charge diffusion constant (or the diffusive current) and electron convection constant (or the convective current) are represented by $\mathcal{D}$ and $\beta$, respectively.

By combining equations (1) and (2), we can obtain the governing equations in GNOR theory given by:

$$\nabla \times \nabla \times E(r,\omega) = (\omega/c)^2 [\varepsilon(\omega) + \xi_{\text{GNOR}} \nabla(\nabla \cdot)] E(r,\omega) \quad (3)$$

Where $\xi_{\text{GNOR}} = \varepsilon_{core}(\omega)[\beta^2 + \mathcal{D}(\gamma - i\omega)]/\omega(\omega+i\gamma)$ is the GNOR nonlocal parameter, $\varepsilon(\omega) = \varepsilon_{core}(\omega) + i\sigma(\omega)/\varepsilon_0\omega$ is the Drude-like permittivity of the metal.

Based on the GNOR wave equation (3), we could further decompose the transverse and longitudinal electric field (local and nonlocal) as follows:

$$(\nabla^2 + k_T^2)\nabla \times E(r,\omega) = 0 \quad (4)$$

$$(\nabla^2 + k_L^2)\nabla \cdot E(r,\omega) = 0 \tag{5}$$

Here, $k_T^2 = (\omega/c)^2 \varepsilon(\omega)$ and $k_L^2 = \frac{\varepsilon(\omega)}{\xi_{GNOR}^2}$ are the wave vectors of the transverse and longitudinal electric fields.

As demonstrated in **Figure 1a**, two particles can be considered indistinguishable if they exhibit equivalent scattering and extinction properties in both the near and far field, regardless of the incident frequencies and angles. To avoid the need for implementing a *k*-dependent permittivity, the elegant local analogue model (LAM) can be utilized to approximate the nonlocal optical response of metal nanoparticles with the fully local optical response of a layered system, wherein the metal surface is coated with thin layers. The dielectric permittivity of the thin layer should satisfy the following condition: [21]

$$\varepsilon_t(\omega) = [\varepsilon(\omega)^{\frac{3}{2}}\varepsilon_b(\omega)\omega(\omega + i\gamma)d]/[\varepsilon(\omega) - \varepsilon_b(\omega)] \cdot \varepsilon_{core}(\omega)[\beta^2 + \mathcal{D}(\gamma - i\omega)] \tag{6}$$

The thickness of the thin layer is represented by $d$, whereas the $\varepsilon_b(\omega)$ denotes the dielectric permittivity of the background. In the case of a particle on a substrate, it is practical to establish a quasi-homogeneous environment by selecting the dielectric permittivity of the background to match that of the substrate $[\varepsilon_s(\omega) = \varepsilon_b(\omega)]$. As long as $d$ is substantially smaller than the metal skin depth, Equation (6) remains valid. A value of d=0.01 nm was used in subsequent simulations for simplicity. Additionally, the response from the bound electrons $\varepsilon_{core}(\omega)$ can be calculated using the measured bulk dielectric functions $[\varepsilon(\omega) = \varepsilon_{exp}(\omega)]$, with the following equation:

$$\varepsilon_{core}(\omega) = \varepsilon_{exp}(\omega) + \omega_p^2/(\omega^2 + i\gamma\omega) \tag{7}$$

Here, $\omega_p$ is the plasma frequency of the metal.

The GNOR-LAM can be easily implemented using the boundary element method (BEM) simulation platform, as shown in **Figure 1b**. To validate the GNOR-LAM, we have conducted simulations of the extinction cross-section of metal spheres under plane wave excitation with identical metal parameters as in the original GNOR paper (see **Figure S1** for detailed results). Our results demonstrate excellent agreement between the simulation outcomes obtained through the GNOR-LAM and those from the original GNOR description. Therefore, we used this method to examine the role of $\mathcal{D}$ and $\beta$ in determining the nonlocal optical response of single gold nanorods. To accomplish this, we fit the Drude model to experimentally obtained material parameters, which yields: $\hbar\omega_p = 8.94$ eV, $\gamma = 0.062$ eV, $\varepsilon_\infty = 9.80$ (**Figure S2**).[24] We then used these parameters to calculate the dielectric permittivity of the thin layer and extract the optical response from the bound electrons using $\varepsilon_{core}(\omega) = \varepsilon_{exp}(\omega) + \omega_p^2/(\omega^2 + i\gamma\omega)$, which was used in the following GNOR-LAM simulation. We have simulated the scattering spectrum of a gold nanorod (12nm × 47 nm, modeled as spherically capped cylinders) under plane wave excitation (**Figure 1b**). Our simulation results show that increasing the value of $\beta$ shifts plasmon resonance energy ($E_{res}$) to higher energies. However, in the absence of $\mathcal{D}$, the plasmon linewidth ($\Gamma$) remains unchanged (**Figure 1c**). Interestingly, both the $E_{res}$ and $\Gamma$ depend on the $\mathcal{D}$. Specifically, for small $\mathcal{D}$ values, no significant change in the $E_{res}$ is observed while the $\Gamma$ experiences notable broadening. In contrast, large $\mathcal{D}$ values lead to a significant blue shift of the $E_{res}$ and broadening of the $\Gamma$. These results suggest that induced charge diffusion effects play a crucial role in determining the nonlocal optical response of plasmonic systems (**Figure 1d**).

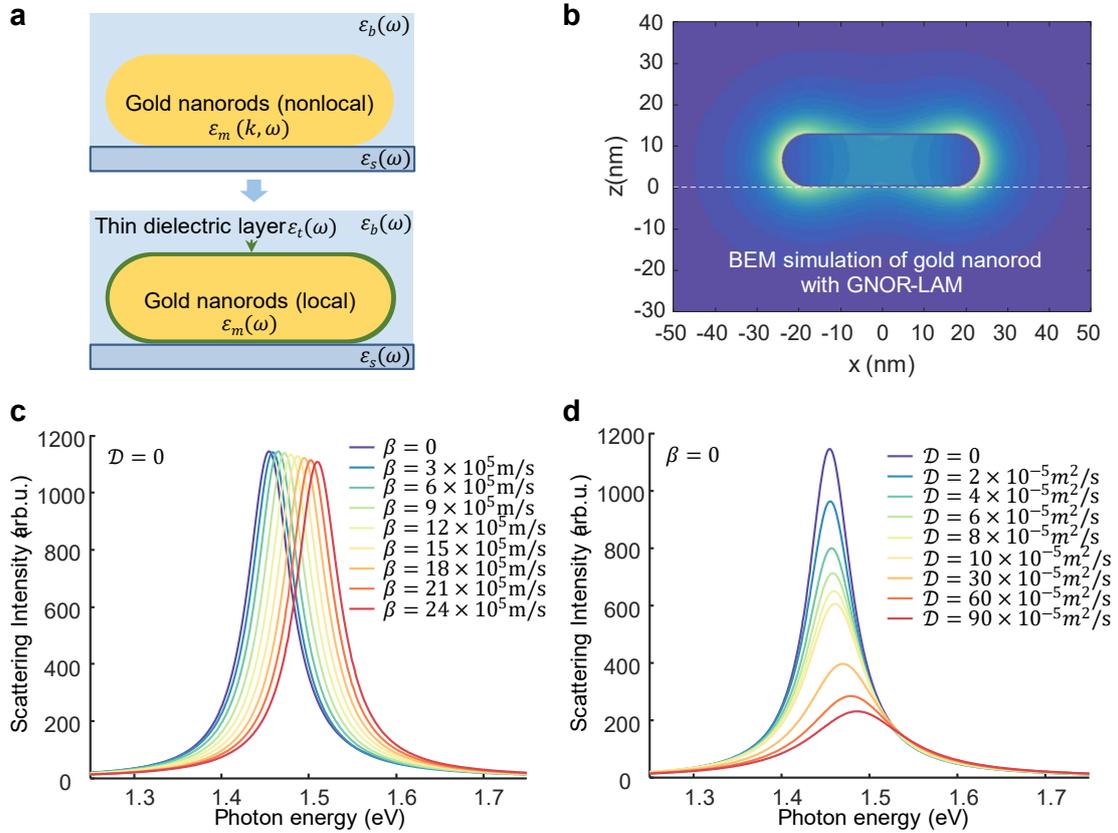

**Figure 1. The generalized nonlocal optical response theory-based local analogue model (GNOR-LAM).** **(a)** A schematics representation of a gold nanorod within the nonlocal description and the local analogue model (LAM) descriptions, where induced charge diffusion and electron convection effects are mapped to a dielectric cover layer. **(b)** The GNOR-LAM can be readily implemented with the boundary element method (BEM) simulation platform. Here, we show the electric field distribution of a gold nanorod (12nm × 47 nm) under plane wave excitation with GNOR-LAM description. **(c)** The scattering spectrum of the same gold nanorods, but with different electron convection constants ($\beta$). The electron convection effects induce a significant blue shift of the plasmon resonance energy. However, the plasmon linewidth remains unchanged in the absence of induced charge diffusion constants ($\mathcal{D}$). **(d)** Both the plasmon resonance energy and linewidth depend on the $\mathcal{D}$. Specifically, for small $\mathcal{D}$ values, no significant change in the plasmon resonance energy is observed, while the plasmon linewidth experiences notable broadening. Conversely, large $\mathcal{D}$ values lead to a significant blue shift of the plasmon resonance energy and broadening of the plasmon linewidth. These results suggest that induced charge diffusion effects play a crucial role in determining the nonlocal optical response of plasmonic systems.

In this work, we used gold nanorods with five different sizes as the model system for our study. There are two reasons for this: first, we could easily synthesize gold nanorods with well-controlled sizes and aspect ratios using established recipes, and second, we could tune their dipole plasmon resonances to the near-infrared region where inter-band damping is negligible. The diameter of the nanoparticles used in our study ranged from 12.51 to 22.54 nm, with aspect ratios ranging from 3.10 to 3.76 (**Figure S3** and **Table S1**). We used transmission electron microscope (TEM) and single-particle darkfield spectroscopy to correlate the mean dimensions to the mean $E_{res}$ and $\Gamma$ in each batch of gold nanorods. To achieve this, we first determined the diameter and length of thousands of nanorods from TEM images. A representative TEM image of nanorods and their corresponding diameter and length are shown in **Figure 2a** and **Figure 2b**. Following this, we measured the scattering spectra of each batch of gold nanorods with our homebuilt dark-field microscope. The obtained scattering spectrum allowed us to determine the $E_{res}$ and $\Gamma$ for every particle by fitting them with the Lorenzian function. A representative scattering spectrum of a single nanorod is shown in **Figure 2c**, which fits well to a Lorentzian function for extracting the $E_{res}$ and $\Gamma$. To obtain better statistics, we recorded the scattering spectra of thousands of nanoparticles using our automated dark-field microscope (The details of the microscope setup description were presented in Supporting Information). Specifically, to measure the scattering spectra of each nanorod, we randomly deposited one batch of nanorods on the Quartz slides and washed the particles thoroughly with a water/ethanol mixture (50%/50%) to remove excess surfactant. After the washing process, we covered the particles with nonpolar mineral oil and coverslips on top of the oil to ensure a homogeneous refractive index ($n \approx 1.47$) around the gold nanorods. Our dark-field microscope automatically identified single nanorods as bright spots and directed a scanning stage so that the scattered light from each one could be spectrally resolved consecutively with a spectrometer. **Figure 2d** shows a plot of the $\Gamma$ versus the $E_{res}$ for one batch of gold nanorods. It should be noted that the $\Gamma$ has negligible dependence on $E_{res}$ in the near-infrared region due to the absence of inter-band damping. We repeated this procedure for all batches of gold nanorods used in this work. Since both $\mathcal{D}$ and $\beta$ are related to Fermi velocity based on the Boltzmann transport equation, it is practical to investigate how they influence $\Gamma$ and $E_{res}$ of gold nanorods with different average distance of electrons to the metal surface (or effective path length of electrons, $l_{eff}$). For gold nanorods, the effective path length of electrons can be calculated from the particle volume $V$ and surface area $S$ according to $l_{eff} = 4V/S$.[25] We approximated the gold nanorods as spherically capped cylinders and used the diameter and length measured by TEM to calculate $l_{eff}$. By correlating the $<E_{res}>$ and $<\Gamma>$ with the $<l_{eff}>$ of the gold nanorods, we were able to analyze the impact of $\mathcal{D}$ and $\beta$ on the optical properties of the nanoparticles.

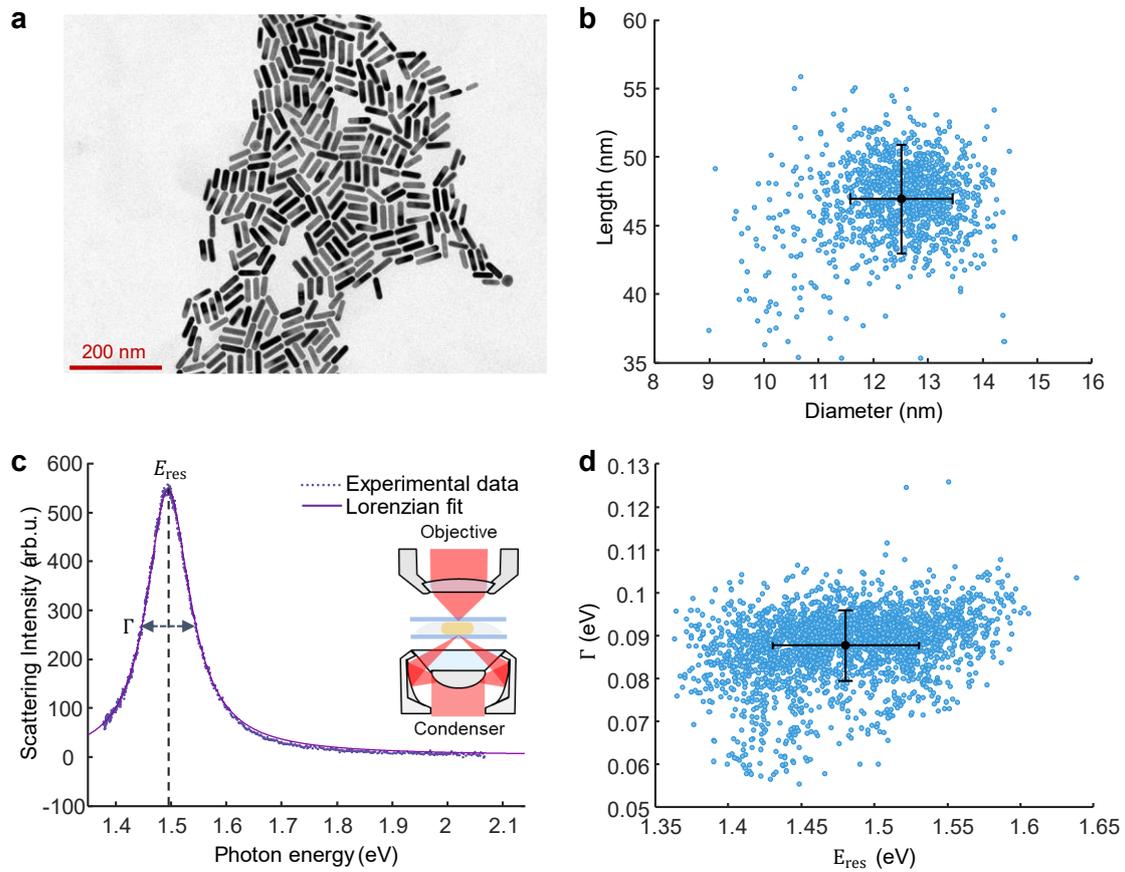

**Figure 2. Correlating mean dimension to the mean plasmon resonance energy and linewidth in the same batch of gold nanorods.** (a) Representative transmission electron microscopy (TEM) image of the gold nanorods used in this work. (b) Thousands of nanorods were analyzed to obtain their diameter and length. Each data point represents a single nanoparticle, which allows us to calculate the mean dimension and effective path length of electrons ($l_{eff}$) of the gold nanorods. (c) The plasmon resonance energy and linewidth of the same batch of gold nanorods were measured using single-particle darkfield scattering spectroscopy. (d) The distribution of plasmon resonance energy and linewidth of the same batch of gold nanorods. This investigation allows us to establish correlations between the physical dimensions and optical properties of the gold nanorods.

In order to quantitatively estimate the values of $\mathcal{D}$ and $\beta$, we first neglect nonlocal effects, and turn to the LRA model to simulate the scattering spectrum of single nanorods on a quartz substrate using the BEM toolbox. Specifically, we simulated the scattering spectra for gold nanorods (with the mean dimensions determined from TEM images) embedded in a medium with a refractive index $n = 1.47$. The gap size between the nanorod and quartz substrate was set at 1.6 nm to mimic the effect of the remaining surfactant.[26] In **Figure 3a**, the purple pentagrams and black dots with error bars (standard deviation) show the simulated $E_{res}$ values obtained from the LRA model, which were compared with the experimentally obtained $<E_{res}>$ values of gold nanorods using single-particle dark-field spectroscopy. The tabulated values for the optical constant of bulk gold and quartz were used for the simulation. However, we found that the traditional LRA model is not sufficient for describing the optical response of plasmonic nanoparticles with small sizes (e.g., $l_{eff} < 20\ nm$). Therefore, we included nonlocal corrections using the GNOR-LAM description to explain our experimental results. Previous studies have shown that $\beta$, which measures the degree of nonlocality, is proportional to the Fermi velocity. In the high-frequency limit ($\omega \gg \gamma$), where longitudinal electromagnetic waves become one-dimensional in nanoparticles, $\beta$ can be described as $\beta = \sqrt{3/5 v_F^2}$.[27] Since $\beta$ only affects the position of E$_{res}$, we have simulated the scattering spectra for gold nanorods with $\beta = \sqrt{3/5 v_F^2} \approx 1.08 \times 10^6 m/s$ (with $v_F \approx 1.39 \times 10^6 m/s$ is the Fermi velocity in gold) and $\mathcal{D}=0$. The simulated E$_{res}$ values (green circles in **Figure 3a**) were compared to the experimentally obtained < E$_{res}$> values, and were found to fit quite well. The small discrepancy can be attributed to the non-perfectly spherocylindrical rod shape of the particle, the remaining surfactant on the particle, and the neglect of $\mathcal{D}$. Next, we have fixed the value of $\beta$ and increased the value of $\mathcal{D}$ from $4 \times 10^{-5} m^2/s$ to $10 \times 10^{-4} m^2/s$. As $\mathcal{D}$ is related to the nonlocal damping, we have obtained an estimated value of $\mathcal{D} \approx 3.1 \times 10^{-4} m^2/s$ by comparing the simulated $\Gamma$ of gold nanorods and experimentally obtained $<\Gamma>$ of gold nanorods (**Figure 3b**). The value of $\mathcal{D}$ obtained in this work is smaller than the previously reported experimental value in superlattice monolayer of gold nanoparticles ($\mathcal{D} \approx 8.8 \times 10^{-4} m^2/s$), and ultrathin metal-dielectric-metal planar structures ($\mathcal{D} \approx 8.0 \times 10^{-4} m^2/s$). A recent report has suggested that the origin of the $\mathcal{D}$ term cannot arise independently from the bulk properties of the plasmonic nanostructures, but rather it has importance at the surface and may be related to surface damping and interface damping effects.[28] Our experimental results indicate the value of $\mathcal{D}$ in a single nanoparticle is smaller than in the plasmon-coupled system. In other words, plasmon coupling may enhance surface damping and interface damping, which play an important role in potential applications using plasmonic nanoparticles for light energy conversion and enhanced Raman spectroscopy. Although the GNOR-LAM description shows an excellent agreement with the experimental results, our simulation describes the particles in an idealized situation. One complication is that the particles are sitting on a substrate, which may have remaining surfactant on one side of the surface and induce additional surface scattering (and possibly interface damping) in our particles. Furthermore, the additional surface scattering (and possibly interface damping) will also induce a higher value of $\mathcal{D}$, which means that the diffusion constant for the clean interface of gold nanorods is anticipated to be even smaller. Interestingly, Kreibig and co-workers proposed many years ago that interface damping (or chemical interface damping) is related to the ability of plasmons to decay by coupling to interfacial electronic states.[29,30] Recently, Foerster and co-workers found that surface damping and interface damping have the same form of phenomenology description, which depends on the average distance of electrons to the surface, the Fermi velocity, and a proportionality constant. This implies that surface damping and interface damping are two aspects of the same process.[25] In the initial GNOR theory, it was consolidated that the value $\mathcal{D}$ is effectively linked to the Kreibig proportionality constant. In this context, our method may offer a way to unify surface scattering damping, chemical interface damping, and Landau damping into induced charge diffusion effects. One direction for future investigation will be to explore how the $\mathcal{D}$ value scales with the surface coverage of molecules or the interfacial states on the particle surface. A detailed investigation of this direction will be left for future investigation.

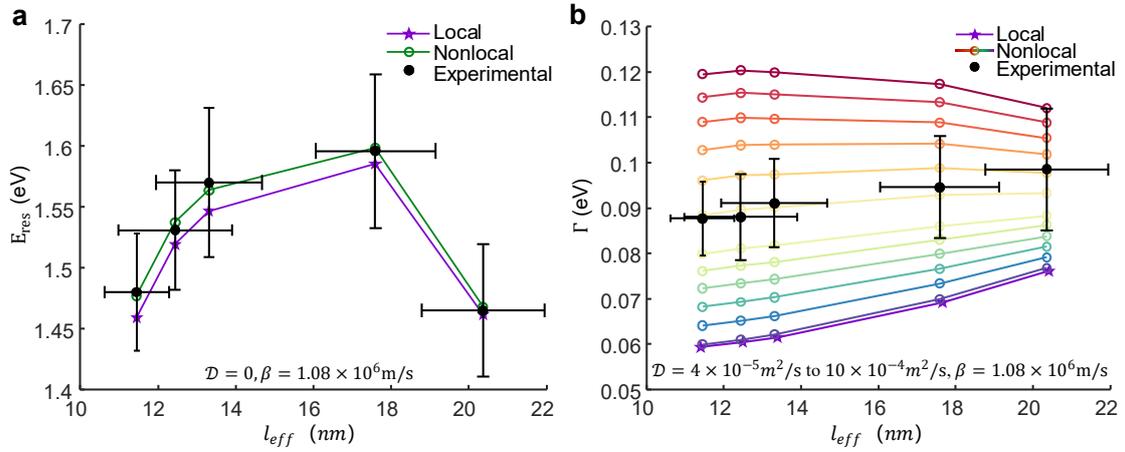

**Figure 3. Quantitatively estimation of $\mathcal{D}$ and $\beta$ values in single gold nanorods.** (a) The traditional local response approximation (LRA) model (purple pentagrams) is insufficient for describing the plasmon resonance energy of plasmonic nanoparticles. Instead, the GNOR-LAM description with $\beta = \sqrt{3/5 v_F^2} \approx 1.08 \times 10^6 m/s$ and $\mathcal{D}=0$ (green circles) matches excellently with experimental obtained $<E_{res}>$ of gold nanorods. (b) To estimate the values of $\mathcal{D}$, we fixed the value of $\beta = 1.08 \times 10^6 m/s$ and increased the value of $\mathcal{D}$ from $4 \times 10^{-5} m^2/s$ to $10 \times 10^{-4} m^2/s$. Our results indicated that increased $\mathcal{D}$ values lead to a significant broadening of the $\Gamma$, especially for the small particles. By comparing the simulated simulated $\Gamma$ of gold nanorods to the experimentally obtained mean $<\Gamma>$ of gold nanorods, we estimated a value of $\mathcal{D} \approx 3.1 \times 10^{-4} m^2/s$.

In summary, we have investigated the nonlocal optical response of single gold nanorods using single-particle dark-field spectroscopy, both experimentally and theoretically. By analyzing the scattering spectra of gold nanorods with different sizes, we have observed the predicted nonlocal optical response of particle plasmons according to the GNOR theory. Furthermore, we have estimated the values of the induced charge diffusion constant and electron convection constant by comparing the simulated and experimentally obtained scattering spectra of gold nanorods. Our results indicate that induced charge diffusion effects in single nanoparticles are less pronounced than those in plasmon-coupled systems. We anticipate that the approach presented in this study will motivate further experimental efforts to explore the induced charge diffusion constant for different dielectric-metal interfaces, which is crucial for the application of particle plasmons in the field of nanophotonics.

## ASSOCIATED CONTENT

**Supporting Information**. Supporting Information contains electron microscopy images and microscope setup description. This material is available free of charge via the Internet at http://pubs.acs.org.

## AUTHOR INFORMATION

**Corresponding Author**

**Weixiang Ye**

*wxy@hainanu.edu.cn

**Author Contributions**

W.Y. conceived, designed, and carried out the research independently.

## ACKNOWLEDGMENT

This research was supported by the National Natural Science Foundation of China (62105229), the Hainan Provincial Natural Science Foundation of China (122RC538), and the Start-up Research Foundation of Hainan University (KYQD(ZZ)21165).

W.Y. would like to express his gratitude to Dr. Carsten Sönnichsen and Dr. Sirin Celiksoy from the University of Mainz for their constructive discussions regarding particle plasmons over the past few years.

## REFERENCES

1. F. J. García de Abajo, R. Sapienza, M. Noginov, et al., "Plasmonic and new plasmonic materials: general discussion," Faraday Discuss. 2015, 178 (0), 9−36.

2. R. F. Oulton, V. J. Sorger, T. Zentgraf, R.-M. Ma, C. Gladden, L. Dai, G. Bartal, X. Zhang, "Plasmon lasers at deep subwavelength scale," Nature 2009, 461 (7264), 629−632.

3. J. B. Pendry, D. Schurig, D. R. Smith, "Controlling electromagnetic fields," Science 2006, 312 (5781), 1780−1782.

4. L. Novotny, N. van Hulst, "Antennas for light," Nat. Photonics 2011, 5 (2), 83−90.

5. X. Hao, M. Yu, R. Xing, C. Wang, W. Ye, "Parallel frequency-domain detection of molecular affinity kinetics by single nanoparticle plasmon sensors," Appl. Phys. Lett. 2018, 121(24), 243703.

6. W. Ye, M. Goetz, S. Celiksoy, L. Tüting, C. Ratzke, J. Prasad, J. Ricken, S. V. Wegner, R. Ahijado-Guzmán, T. Hugel, C. Soennichsen, "Conformational Dynamics of a Single Protein Monitored for 24 h at Video Rate," Nano Lett. 2018, 18(10), 6633-6637.

7. W. Ye, S. Celiksoy, A. Jakab, A. Khmelinskaia, T. Heermann, A. Raso, S. V. Wegner, G. Rivas, P. Schwille, R. Ahijado-Guzmán, C. Soennichsen, "Plasmonic Nanosensors Reveal a Height Dependence of MinDE Protein Oscillations on Membrane Features," J. Am. Chem. Soc. 2018, 140(51), 17901-17906.

8. A. Al-Zubeidi, L. A. McCarthy, A. Rafiei-Miandashti, T. S. Heiderscheit, S. Link, "Single-particle scattering spectroscopy: fundamentals and applications," Nanophotonics 2021, 10, 1621-1655.

9. W. Ye, K. Krüger, A. Sánchez-Iglesias, I. García, X. Jia, J. Sutter, S. Celiksoy, B. Foerster, L. M. Liz-Marzán, R. Ahijado-Guzmán, C. Sönnichsen, "Metal-Semiconductor Hybrid Dimer Nanostructures for Low-Power and High-Frequency Modulation of Near-Infrared Light," Chem. Mater. 2020, 32, 1650-1656.

10. B. Foerster, J. Rutten, H. Pham, S. Link, C. Sönnichsen, "Shaping the Plasmonic Response: Importance of the Chemical Interface Damping," J. Phys. Chem. C 2018, 122, 19116–19123.

11. A. V. Uskov, I. E. Protsenko, N. A. Mortensen, E. P.O'Reilly, "Broadening of plasmonic resonances due to electron collisions with nanoparticle boundary: quantum-mechanical consideration," Plasmonics 2014, 9, 185−192.

12. N. A. Mortensen, S. Raza, M. Wubs, T. Søndergaard, S. I. Bozhevolnyi, "A generalized nonlocal optical response theory for plasmonic nanostructures," Nat. Commun. 2014, 5, 3809.

13. G. Toscano, J. Straubel, A. Kwiatkowski, C. Rockstuhl, F. Evers, H. Xu, N. A. Mortensen, M. Wubs, "Non-local optical response of metallic nanowires probed by angle-resolved cathodoluminescence spectroscopy," Nat. Commun. 2015, 6, 7132.

14. J. A. Scholl, A. L. Koh, J. A. Dionne, "Quantum plasmon resonances of individual metallic nanoparticles," Nature 2012, 483 (7390), 421−U68.

15. S. Raza, N. Stenger, S. Kadkhodazadeh, S. V. Fischer, N. Kostesha, A.-P. Jauho, A. Burrows, M. Wubs, N. A. Mortensen, "Blueshift of the surface plasmon resonance in silver nanoparticles: substrate effects," Nanophotonics 2013, 2 (2), 131−138.

16. C. Ciracì, R. T. Hill, J. J. Mock, Y. Urzhumov, A. I. Fernández-Domínguez, S. A. Maier, J. B. Pendry, A. Chilkoti, D. R. Smith, "Probing the Ultimate Limits of Plasmonic Enhancement," Science 2012, 337(6098), 1072–1074.

17. H. Shen et al., "Optical observation of plasmonic nonlocal effects in a 2D superlattice of ultrasmall gold nanoparticles," Nano Lett. 2017, 17, 2234–2239.

18. S. Boroviks, Z.-H. Lin, V. A. Zenin, M. Ziegler, A. Dellith, P. A. D. Gonçalves, C. Wolff, S. I. Bozhevolnyi, J.-S. Huang, N. A. Mortensen, "Extremely confined gap plasmon modes: when nonlocality matters," Nat. Commun. 2022, 13, 3105.

19. C. Sönnichsen, T. Franzl, T. Wilk, G. von Plessen, J. Feldmann, O. Wilson, P. Mulvaney, "Drastic reduction of plasmon damping in gold nanorods," Phys. Rev. Lett. 2002, 88, 077402.

20. U. Hohenester, A. Trügler, "MNPBEM - A Matlab toolbox for the simulation of plasmonic nanoparticles," Comp. Phys. Commun. 2012, 183, 370.

21. Y. Luo, A. I. Fernandez-Dominguez, A. Wiener, S. A. Maier, J. B. Pendry, "Surface Plasmons and Nonlocality: A Simple Model," Phys. Rev. Lett. 2013, 111, 093901.

22. T. Y. Dong, K. Yin, X. K. Gao, X. K. Ma, "Generalized local analogue model for nonlocal plasmonic nanostructures based on multiple-fluid hydrodynamic framework," J. Phys. D Appl. Phys. 2020, 53, 295105.

23. N. A. Mortensen, "Mesoscopic electrodynamics at metal surfaces - From quantum-corrected hydrodynamics to microscopic surface-response formalism," Nanophotonics 2021, 10(10).

24. G. Rosenblatt, B. Simkhovich, G. Bartal, M. Orenstein, "Nonmodal Plasmonics: Controlling the Forced Optical Response of Nanostructures," Phys. Rev. X. 2020, 10, 011071.

25. B. Foerster, A. Joplin, K. Kaefer, S. Celiksoy, S. Link, C. Sönnichsen, "Chemical Interface Damping Depends on Electrons Reaching the Surface," ACS Nano 2017, 11 (3), pp 2886–2893.

26. S. Meena, S. Celiksoy, P. Schäfer, A. Henkel, C. Sönnichsen, M. Sulpizi, "The role of hydrophobicity and hydration in protein-ligand binding studied by computer simulations and single molecule fluorescence spectroscopy," Phys. Chem. Chem. Phys. 2016, 18, 13246-13254.

27. A. Moreau, C. Ciraci, D. R. Smith, "Impact of nonlocal response on metal-dielectric multilayers and optical patch antennas," Phys. Rev. B 2013, 87, 045401.

28. M.K. Svendsen, C. Wolff, A.-P. Jauho, N.A. Mortensen, C. Tserkezis, "Role of diffusive surface scattering in nonlocal plasmonics," Journal of Physics: Condensed Matter 32, 395702 (2020).

29. U. Kreibig, C. Fragstein, "The limitation of electron mean free path in small silver particles," Z. Physik. 224, 307–323 (1969).

30. U. Kreibig, M. Vollmer, "Optical Properties of Metal Clusters," Springer-Verlag, 1995.

31. B. Foerster, M. Hartelt, S. S. E. Collins, M. Aeschlimann, S. Link, C. Sonnichsen, "Interfacial states cause equal decay of plasmons and hot electrons at gold-metal oxide interfaces," Nano Lett. 20, 3338-3343 (2020).

# Supporting Information for

## Nonlocal Optical Response of Particle Plasmons in Single Gold Nanorods

*Weixiang Ye[1], \**

[1] Center for Theoretical Physics and School of Science, Hainan University, Haikou 570228, China

*Corresponding author: wxy@hainanu.edu.cn

**This PDF file includes:**

    Materials and Methods
    Supplementary Text
    Figures S1 to S3
    Table S1

## MATERIALS AND METHODS

**Materials.** In this study, we utilized the following chemical reagents: Gold chlorate trihydrate (HAuCl$_4$·3H$_2$O, ≥ 99.9 %), sodium borohydrate (NaBH$_4$, 99.99 %), sodium chloride (NaCl, > 99.0 %), Hexadecyltrimethylammonium chloride (CTAC, > 98.0%), hydrogen chloride (HCl, 37 wt. % in water), Hexadecyltrimethylammonium bromide (CTAB, > 98.0%), and L-ascorbic Acid (BioUltra, ≥ 99.5%) were obtained from Sigma Aldrich. Silver nitrate (AgNO$_3$, ≥ 99.9 %) was purchased from Carl Roth, while Sodium oleate (NaOL, > 97 %) was bought from TCI. Quartz microscope slides and coverslips were purchased from Fisher Scientific. Deionized water from a Merck Millipore system (>18 MΩ, MilliQ) was used for all experiments.

**Microscope setup.** Single particle scattering spectra were acquired using a Zeiss microscope (Axio Imager.M2m) that was equipped with a Plan–Apochromat objective (63/1.4 Oil Iris M27), a Piezostage PI542 and a Z-Piezo PIFOC-721. We used a spectrometer (ImSpector V10E) and a CMOS camera (Hamamatsu Orca Flash 4.0 V3) to obtain the spectra. A home-written MATLAB software package was used for the data acquisition. The nanorods were automatically centered at the slit of the spectrometer as well as focused by optimizing the intensity.

**Synthesis and characterization of gold nanorods.** We synthesized five batches of gold nanorods with varying diameters and aspect ratios using a two-step seeded-growth process outlined in the literature (details parameters specified in **Table S1**) [1]. We characterized the nanorods by transmission electron microscopy (TEM) using a Tecnai G2 Spirit FEI electron microscope, operating at an acceleration voltage of 120 keV. Representative images are provided in Figure S3. We determined the diameter (D) and length (L) of thousands of nanorods from these images by modeling the gold nanorods as spherically capped cylinders, and the resulting $<D>, <L>$, and $<l_{eff}>$ are reported in Table S2.

## REFERENCES

1. Ye, X.; Gao, Y.; Chen, J.; Reifsnyder, D. C.; Zheng, C.; Murray, C. B. Seeded Growth of Monodisperse Gold Nanorods Using Bromide-Free Surfactant Mixtures. **Nano Lett**. 2013, 13, 2163–2171.
2. Rosenblatt, G.; Simkhovich, B.; Bartal, G.; Orenstein, M. Nonmodal Plasmonics: Controlling the Forced Optical Response of Nanostructures. **Phys. Rev. X**. 2020, 10, 011071.
3. Mortensen, N. A.; Raza, S.; Wubs, M.; Søndergaard, T.; Bozhevolnyi, S. I. A Generalized Non-local Optical Response Theory for Plasmonic Nanostructures. **Nat. Commun.** 2014, 5, 3809.

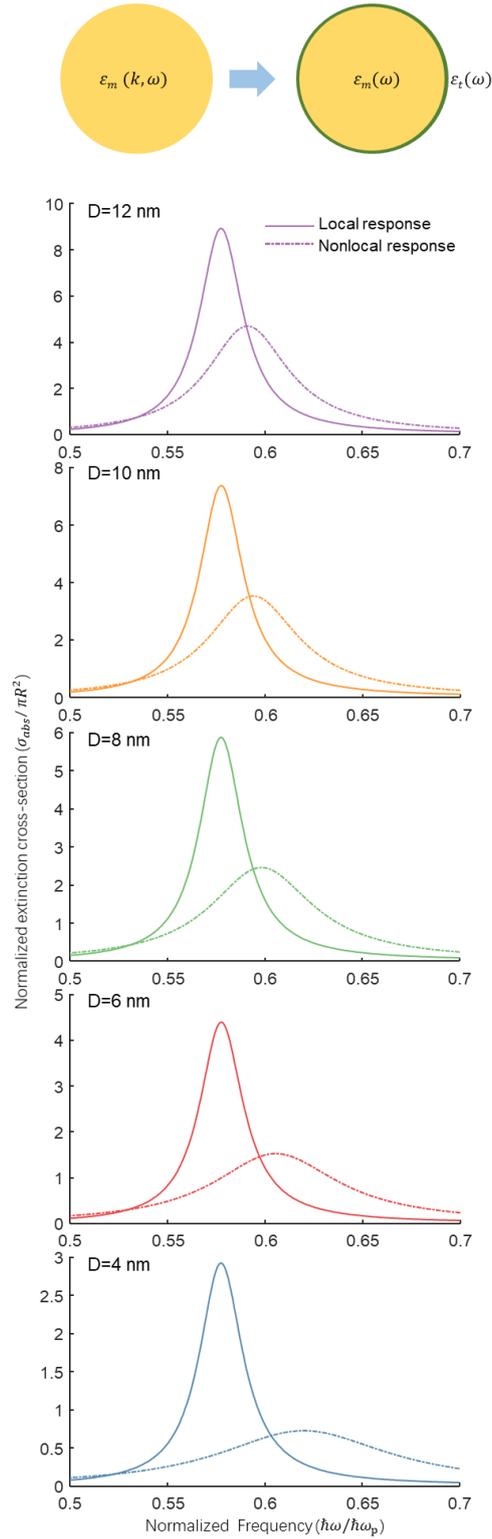

**Figure S1. Extinction cross-section of metal sphere under plane wave excitation using generalized nonlocal optical response theory-based local analogue model (GNOR-LAM) description.** To ensure direct comparability, we used the identical metal parameters in Ref. [2]: $\hbar\omega_p$ = 5.89 eV, $\gamma$ = 0.16 eV, $\beta = 0.81 \times 10^6$ m/s and $\mathcal{D} = 2.04 \times 10^{-5} m^2/s$. Our results demonstrate that the simulation outcomes from the GNOR-LAM description are in excellent agreement with those from the original GNOR description.

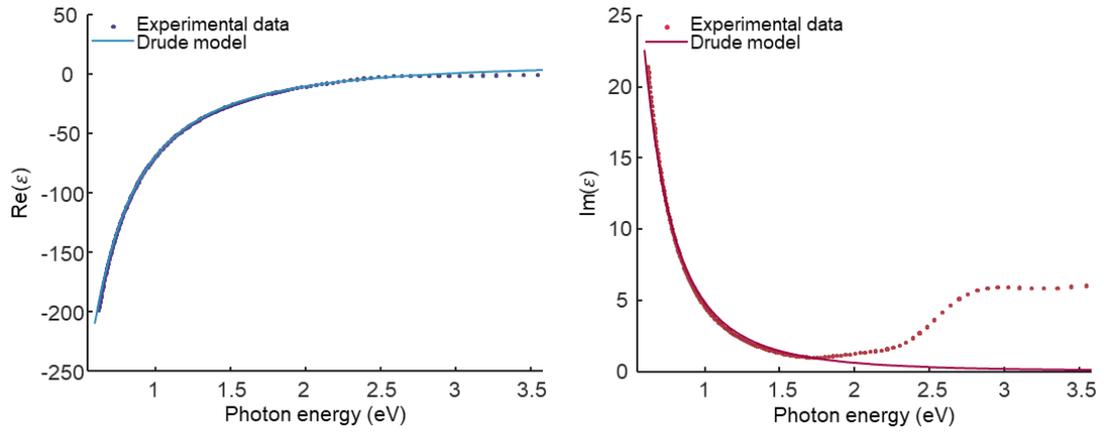

**Figure S2**. **Experimental data and the Drude model for the real and imaginary parts of the dielectric function of bulk gold are presented.** The experimental data (dots) were obtained from ellipsometry measurements in Ref.[3], while the Drude model (line) was calculated using the fitting parameters: $\hbar\omega_p = 8.94$ eV, $\gamma = 0.062$ eV, $\varepsilon_\infty = 9.80$.

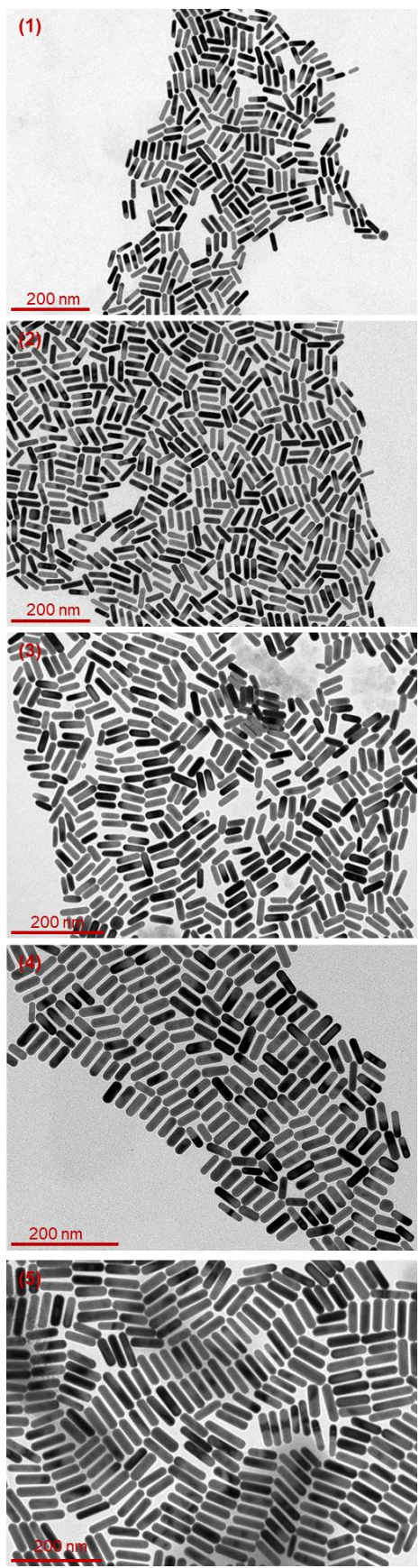

**Figure S3**. **Representative transmission electron microscopy (TEM) images of the 5 batches of gold nanorods used in this work.** The scale bar is set at 200 nm and the particles have been arranged in order of diameter.

| Sample | CTAC (g) | AgNO3 (mL) | Seed (mL) | HCl (mL) | NaOL (g) |
|---|---|---|---|---|---|
| #1 | 6.15 | 12.0 | 1.6 | 3.0 | 1.543 |
| #2 | 5.0 | 12.0 | 1.2 | 3.0 | 1.543 |
| #3 | 6.15 | 12.0 | 1.2 | 3.0 | 1.543 |
| #4 | 6.15 | 12.0 | 0.8 | 3.0 | 1.543 |
| #5 | 5.44 | 12.5 | 1.2 | 4.0 | 1.543 |

| Sample | Mean Diameter (nm) | Mean Length (nm) | Mean aspect ratio | Mean effective path length (nm) |
|---|---|---|---|---|
| #1 | 12.51±1.15 | 46.95±3.96 | 3.76±0.43 | 11.45±0.83 |
| #2 | 13.82±1.95 | 47.56±3.71 | 3.47±0.41 | 12.44±1.46 |
| #3 | 14.88±1.85 | 49.41±3.53 | 3.34±0.48 | 13.31±1.37 |
| #4 | 19.80±1.68 | 61.13±6.71 | 3.10±0.60 | 17.59±1.54 |
| #5 | 22.54±2.06 | 79.59±6.89 | 3.54±0.52 | 20.36±1.58 |

**Table S1**. **Synthesis details and dimensions of gold nanorods**. This table summarizes the dimensions of gold nanorods, determined by analyzing transmission electron microscopy (TEM) images, as well as the quantities of chemicals used for their synthesis.